\documentclass[twocolumn,showpacs,preprintnumbers,amsmath,amssymb,elsevier]{revtex4}

\usepackage{graphicx}
\usepackage{float}
\usepackage{bm}
\usepackage{color}

\begin{document}

\title{Quantum interference in a Cooper pair splitter: the three sites model}

\author{Fernando Dom\'inguez and Alfredo Levy Yeyati}
\affiliation{Departamento de F\'isica Te\'orica de la Materia Condensada,
        Condensed Matter Physics Center (IFIMAC), \\ and Instituto Nicol\'as
        Cabrera, Universidad Aut\'onoma de Madrid, E-28049 Madrid, Spain}

\date{\today}

\begin{abstract}
New generation of Cooper pair splitters defined on hybrid nanostructures are devices 
with high tunable coupling parameters. 
Transport measurements through these devices revealed clear signatures
of interference effects and motivated us 
to introduce a new model, called the 3-sites model. 
These devices provide an ideal playground to tune the Cooper pair splitting (CPS) 
efficency on demand, and displays 
a rich variety of physical phenomena. 
In the present work we analyze theoretically the conductance of the 3-sites model in the linear and non-linear 
regimes and characterize the most representative features 
that arise by the interplay of the different model parameters. 
In the linear regime we find that the local processes typically 
exhibit Fano-shape resonances, while the CPS contribution exhibits Lorentzian-shapes. 
Remarkably, we find that under certain conditions, the transport is blocked by the 
presence of a dark state.
In the non-linear regime we established a hierarchy of the model parameters to obtain 
the conditions for optimal efficency.
\end{abstract}

\maketitle

\section{Introduction}

Producing and detecting entangled electronic states in nanoscale circuits is a prominent goal 
which is still under development.
Among his many contributions to the field of quantum transport Markus B\"uttiker 
participated in the intense search for manifestations of quantum entanglement 
in transport experiments which took place at the beginning of 2000 
[\onlinecite{Burkard2000a, Deutscher2000a, Recher2001a, Lesovik2001a, Bena2002a, Samuelsson2002a,
Chtchelkatchev2002a,Beenakker2003a,Saraga2004a,Lebedev2005a, Sanchez2008c}]. 
Markus and his group extensively explored the characterization of the degree of entanglement by
cross correlations of the electrical current 
[\onlinecite{Samuelsson2002a, Samuelsson2003a, Samuelsson2004a}]. Within the early theoretical 
proposals for producing entanglement in transport, the ones based on superconducting 
correlations [\onlinecite{Recher2001a, Lesovik2001a ,Bena2002a}] have adquired 
a particular relevance with the first experimental realizations of CPS in semiconducting 
nanowires [\onlinecite{Hofstetter2009a, Hofstetter2011a, Fulop2014a}] 
and carbon nanotubes [\onlinecite{Herrmann2010a, Schindele2012a, Schindele2014a}]. 
These experiments, although providing evidence of CPS through conductance measurements, 
have not yet achieved the goal of demonstrating entanglement.
In connection to these developments there have also appeared 
some proposals for detecting entanglement from conductance measurements
[\onlinecite{Braunecker2013a, Schroer2014a}].

Ideally, the basic 
mechanism for enforcing CPS in these devices is the presence of a large intradot Coulomb repulsion 
and a large gap for quasiparticle excitations in the superconductor.  In actual devices, however, 
geometrical and/or parameters constraints complicate the analysis but,  at the same time, give rise 
to interesting new phenomena. In this sense, while in an ideal CPS device it is assumed that Cooper 
pairs are injected at a slow rate and extracted at a faster rate in order to avoid overlap between 
subsequent pairs, in an actual device the injected pairs can dwell within the device 
and interference effects due to the superposition of different electronic paths can emerge. In 
fact, as shown first in the experiments of Ref.~[\onlinecite{Yacobi1995a}] for QDs inserted in a 
Aharonov-Bohm ring and analyzed theoretically by Markus B\"uttiker and co-authors 
[\onlinecite{Levy1995b,Taniguchi1999a,Levy2000a}], the phase of the transmission through a QD can be well 
defined even when the system is deep the Coulomb blockade (CB) regime, thus leading to interference effects. 
Such effects have been reported in a recent experiment on CPS devices based on InAs nanowires
coupled to a Nb superconducting lead [\onlinecite{Fulop2015a}]. These experiments have motivated the model
that we analyze in detail within the present work. 

Another motivation for our work is the extension of the CPS analysis to the finite bias regime. While
most theoretical analysis and measurements concerning CPS have been restricted to the
linear regime where the applied bias voltage is negligible compared to all energy scales (temperature,
charging energy and the superconducting gap), the applied bias to each lead is an additional parameter
to play with, which can be easily controlled experimentally. It has also been argued [\onlinecite{Recher2001a}] that
a finite bias voltage together with an antisymmetric detuning of the dot resonances can be used to
increase the CPS efficiency by enhancing the non-local with respect to the local processes. 
Again, the validity of these arguments for the actual experimental geometries and parameter regimes
should be tested. 

The rest of the manuscript is organized as follows: 
in section II we present the 3-sites model used to describe 
the interference effects in a Cooper pair splitter (see Ref.~[\onlinecite{Fulop2015a}]). 
Then, in section III, we introduce the Keldysh formalism used to calculate the conductance.
Furthermore, we explain the self-consistent approach used here to calculate the current 
in the presence of Coulomb interactions. 
In section IV we present the conductance results in the linear regime,
and we analyze the evolution of the conductance profiles varying the rest of the model parameters in two different limits: 
one in which the central system is fully hybridized and another in which it is partially hybridized.
Finally in section V, we show some representative conductance calculations in the non-linear regime. 
We focus our attention on the case in which the energy levels of the dots are tuned symmetrically and antisymmetrically. 
Furthermore, we establish a hierarchy of the 3-sites model parameters to  
to obtain the optimal conditions to enhance the CPS efficency.

\section{Description of the 3-sites model}

\begin{figure}[t]
\centering
\includegraphics[width=3.0in,clip]{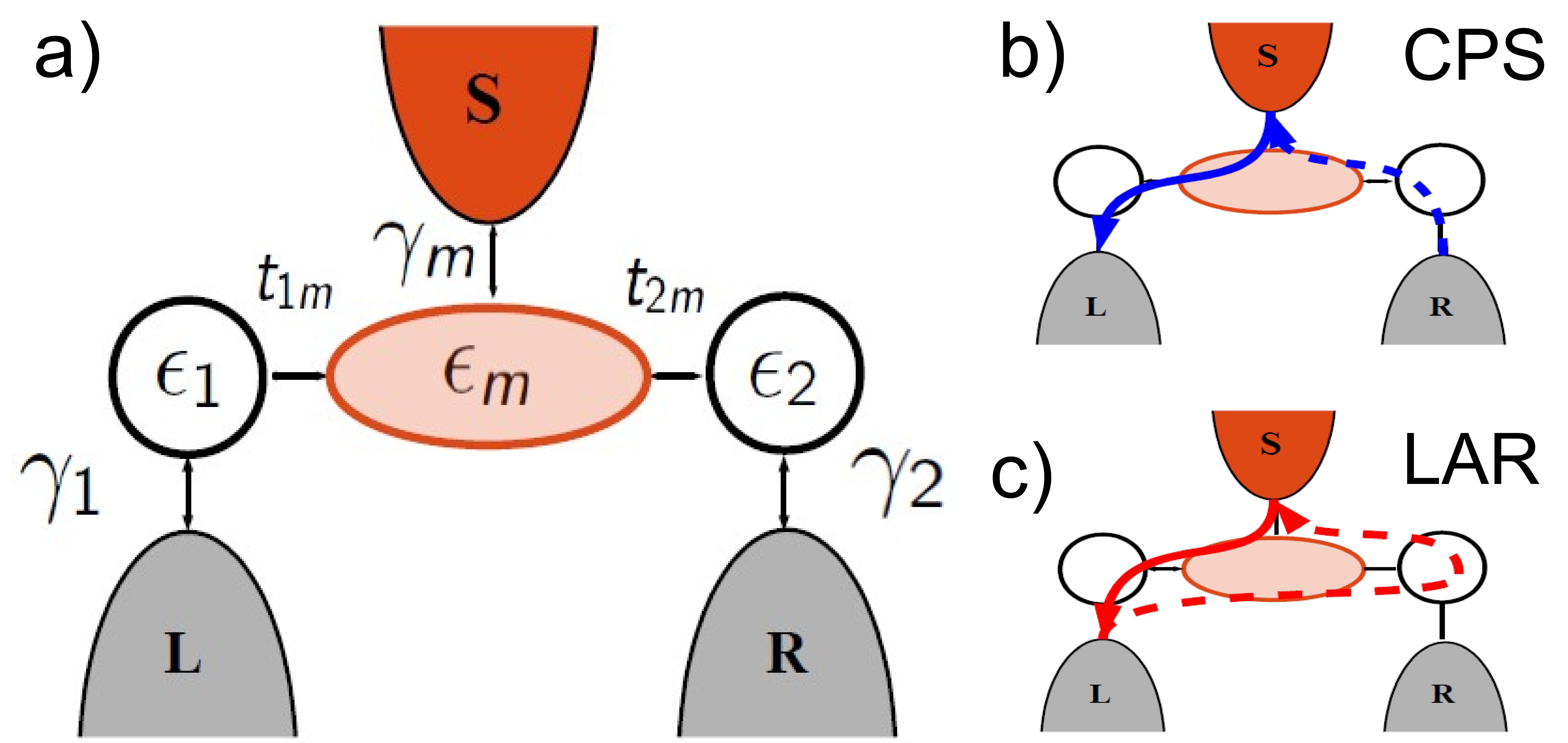}
\caption{Panel~a: We show the schematic representation of the 3-sites model, with the model parameters. 
Panels~b and~c show the electronic (solid line) and the hole (dashed line) paths that the Cooper pairs 
may take depending on the process, CPS and LAR, respectively.}
\label{Fig1}
\end{figure}

In the recent experiments of Ref.~[\onlinecite{Fulop2015a}] the conductance 
through the QDs exhibited Fano-like resonances when operated in the Cooper pair splitting mode.
A qualitative description of the experimental results require to go beyond previous incoherent 
models with independent transport mechanisms only coupled by 
the QD dynamics [\onlinecite{Fulop2014a,Schindele2012a}] or the simpler coherent two-dot models 
considered in Refs.~[\onlinecite{Herrmann2010a,Chevalier2011a,Eldridge2010a}].  
The 3-sites model introduced in Ref.~[\onlinecite{Fulop2015a}] is schematically depicted in Fig.~\ref{Fig1}.

The system can be decomposed in two parts; a coherent central region and the electronic reservoirs.
The coherent part is given by three discrete spin levels coupled coherently.
We will assume that the part of the wire that separates the quatum dots can be 
effectively described by a single discrete level. This approximation can be done 
when the energy separation between the levels of the central part ($\delta \epsilon$) is much higher 
than the coherent tunnel between the central part and the dots, i.e.~$\delta \epsilon\gg t_{im}$.
We will assume that the Coulomb interaction in the central part is negligible because 
it is screened by the nearby superconducting electrode. Thus, we can write the central part by the Hamiltonian
\begin{align}
H_{3s}=\sum_{i,\sigma} \epsilon_{i,\sigma} \hat{n}_{i,\sigma}+U_i \hat{n}_{i\uparrow}\hat{n}_{i\downarrow} + \sum_{i\neq m,\sigma} t_{im} d^\dagger_{i,\sigma} d_{m,\sigma}+\mathrm{h.c.}
\label{eq-H3s}
\end{align}
with $i=1,m,2$, and $d_{i,\sigma}^{(\dagger)}$ destroys (creates) an electron with $\sigma$-spin in the site {\it i}, 
$\hat{n}_{i,\sigma}$ is the number operator, 
$\epsilon_i$ are the site energies, $t_{im}$ are the tunnel coupling amplitudes between the 
site {\it i} and the central site, and $U_i$ are the Coulomb interaction constants. As we mentioned above we set $U_m=0$.
It has been shown experimentally [\onlinecite{Fulop2014a, Fulop2015a}] 
that the energies $\epsilon_{i,\sigma}$ and the tunnel couplings 
$t_{i,m}$ can be tunned by several nearby gate electrodes.
We describe the normal leads $l=1,2$ using non-interacting (normal) Fermi
liquids, described by the Hamiltonian 
\begin{align}
H_{lead-l}=\sum_{{\bf k}\sigma}\epsilon_{\bf k}a_{l{\bf k}\sigma}^{\dagger}a_{l{\bf k}\sigma},
\end{align}
where $a_{l{\bf k}\sigma}^{(\dagger)}$ is the annihilation (creation) operator of an 
electron in the {\it l}-lead.
On the other hand, the superconducting lead is described by the
BCS-Hamiltonian
\begin{align}
H_{S}=\sum_{{\bf k},\sigma} \xi_k c_{{\bf k}\sigma}^\dagger c_{{\bf k}\sigma} -\sum_{\bf k} \left(\Delta_k c_{{\bf k}\uparrow}^\dagger c_{-{\bf k}\downarrow}^\dagger+\mathrm{h.c.}\right),
\end{align}
where $c_{{\bf k},\sigma}^\dagger$ creates a fermion with k momentum and spin $\sigma=\uparrow,\downarrow$. 
The coupling to the leads is given by
\begin{align}
H_\tau = 
\sum_{i,{\bf k},\sigma} \left( 
t_i  d_{i,\sigma}^{\dagger} a_{i,{\bf k},\sigma} + 
 t_{m} d_{i,\sigma}^{\dagger} c_{{\bf k},\sigma} + \mathrm{h.c.}\right).
\label{Eq.coupling}
\end{align}
Here, tunneling from the
dot to the state ${\bf k}$ 
in the lead is described by the tunel amplitude $t_{1,m,2}$.
We assume that the ${\bf k}$-dependence of  the tunel amplitudes 
can be neglected.
These tunel amplitudes lead to the tunel rates defined by $\gamma_i=\pi \rho_i t_i^2$, being $\rho_i$ the density of 
states of the {\it ith}-lead. 

\section{Transport properties}

The mean current through the {\it j}-lead is defined as
\begin{align}
I_j={\it i}\frac{e}{\hbar} \sum_{k,\sigma} t_j
\left( 	\langle a_{k,j,\sigma}^\dagger d_{j,\sigma} \rangle 
-\langle d_{j,\sigma}^\dagger a_{k,j,\sigma} \rangle  \right).
\end{align}
In order to calculate the current, it is
convenient to express this quantity in terms of 
the Keldysh Green's functions (GFs)
\begin{align}
I_j=\frac{e}{\hbar}\int \frac{d\omega}{2\pi}
\textrm{Tr} 
\left \{   
\hat{\tau}_j
\left(
\widehat{G}_{DL}^{+-}(\omega)-\widehat{G}_{LD}^{+-}(\omega)
\right)
\sigma_{z}
\right\},
\label{eqcurr1}
\end{align}
where $\widehat{\tau_j}=t_{j}\sigma_z P_j$, and $P_j$ projects on the {\it j}-subspace, with $j=1,m,2$.
In this expression $\widehat{G}^{\alpha \beta}_{DL}(\omega)$ with $\alpha,\beta=+,-$, is the Fourier transform of the 
Keldysh matrix GF 
$\widehat{G}^{\alpha \beta}_{DL}(t,t')=-{\it i} \langle T_c  \psi(t_{\alpha}) \Lambda^{\dagger}(t'_{\beta})\rangle$ 
and $\widehat{G}^{\alpha \beta}_{LD}(t,t')=-{\it i} \langle T_c  \Lambda(t_{\alpha}) \psi^{\dagger}(t'_\beta)\rangle$ 
and $T_c$ indicates time ordering on the Keldysh contour. 
The GFs are defined using the generalized Nambu spinors
\begin{align}
&\psi^\dagger(t)=\begin{bmatrix}
d_{1\uparrow}^\dagger& d_{m\uparrow}^\dagger & d_{2\uparrow}^\dagger &  d_{1\downarrow}   & d_{m\downarrow}   & d_{2\downarrow} 
\end{bmatrix}_t, \text{and}
\label{eqns}\\
&\Lambda^\dagger(t)=\sum_{k}\begin{bmatrix}
 a_{1\uparrow}^\dagger& c_{k,\uparrow}^\dagger & a_{2,k,\uparrow}^\dagger &  a_{1,k,\downarrow}   & c_{k,\downarrow}   & a_{2,k,\downarrow} 
\end{bmatrix}_t.
\end{align}
We shall assume that spin symmetry is preserved, which allows us 
to use a 6x6 instead of a 12x12 description.

Using the Dyson equation, the Langreth theorem and some standard manipulations we 
rexpress Eq.~\eqref{eqcurr1} in terms of the GFs of the coherent part, that is
\begin{align}
I_j
&=\frac{e}{h}\int d\omega   \textrm{Tr} 
\{   
(
\widehat{G}_{D}^{r} \widehat{\Sigma}^{+-}_j-\widehat{\Sigma}^{+-}_j \widehat{G}_{D}^{a}
\nonumber \\ 
&+
\widehat{G}_{D}^{+-} \widehat{\Sigma}^{a}_j-\widehat{\Sigma}^{r}_j \widehat{G}_{D}^{+-}
)\sigma_z\},
\label{eqcurr}
\end{align}
where 
\begin{align}
\widehat{\Sigma}_j^{\alpha \beta}=\widehat{\tau}_j \hat{g}_{L}^{\alpha \beta}\widehat{\tau}_j,
\end{align}
being $\hat{g}_{L}^{\alpha \beta}$ the Fourier transform of the uncoupled leads
GFs $\hat{g}^{\alpha \beta}_{L}(t,t')=-{\it i} \langle T_c \Lambda(t_{\alpha}) \Lambda^{\dagger}(t'_\beta)\rangle_0$, and
$\widehat{G}^{\alpha \beta}_{D}(t,t')=-{\it i} \langle T_c \psi(t_{\alpha}) \psi^{\dagger}(t'_\beta)\rangle$ 
corresponds to the fully coupled GFs on the central regions.

In the presence of Coulomb interactions 
we need to approximate the GF evaluation.
The approximation that we use consists in starting from the exact 
GFs for the uncoupled central region described by Eq.~\eqref{eq-H3s}, and 
then introducing the coupling to the leads by means of Dyson's equation .
For instance, in the case of the retarded and advanced components, 
\begin{align}
\widehat{G}_D^{r/a} \approx ((\hat{g}_0^{r/a})^{-1}- \widehat{\Sigma}_0^{r/a})^{-1},
\end{align}
where $\widehat{\Sigma}^{r/a}_0$ is the non-interacting self-energy matrix
and $\hat{g}_0^{r/a}$ is the uncoupled to the leads interacting GF (see App.~A).
In addition, for evaluating Eq.~\eqref{eqcurr}, we need $G^{+-}_D$, which is given by 
\begin{align}
\widehat{G}_{D}^{+-}=\widehat{G}_{D}^r\widehat{\Sigma}^{+-} \widehat{G}_{D}^a 
+ (1+\widehat{G}^r_{D}\widehat{\Sigma}^r) \hat{g}_D^{+-}(1+\widehat{\Sigma}^a \widehat{G}^a_{D}),
\label{eqgfgl}
\end{align}
where $\hat{g}^{+-}_D$ is the lesser GF of the isolated central part.
When $\widehat{G}_{D}^{r,a}$ and $\widehat{\Sigma}$ are exact, the second 
term in Eq.~\eqref{eqgfgl} can be set to zero 
because one can always assume an initially empty state. 
However, as we have explained above in the presence of the Coulomb interactions 
one needs to approximate $\widehat{\Sigma}^{r/a}$, and consequently some spurious effects may
arise. 
In particular, it may lead to a violation of current 
conservation, i.e.~$I_1+I_2+I_3\neq 0$.
In order to compensate this artificial effect,
we keep the second term of Eq.~\eqref{eqgfgl}, with
\begin{align}
\hat{g}^{+-}_{D}(\mu_c)=2{\it i} \hat{f}(\mu_c) \text{Im}\{\hat{g}_D^{a}(\omega)\},
\end{align}
where $\widehat{f}$ is a diagonal matrix containing the electron or hole
Fermi functions with chemical potential $\mu_c$.
We tune $\mu_c$ in order that the total current is conserved.
Thus, the inclusion of the second term in Eq.~\eqref{eqgfgl} allows us to obtain a current conserving approximation.

All the calculations are done setting the chemical potential of the superconductor to zero, and 
using a symmetric bias for the normal leads, i.e.~$\mu=\mu_L=\mu_R$, which corresponds to the Cooper pair splitting mode.
We calculate the differential conductance from
$G_j=e \partial I_{j}/\partial \delta U $, using a differential voltage
$\delta U$ applied symmetrically to the normal leads, i.e.~$\mu=\mu_L+\delta U=\mu_R+\delta U$.

It is interesting to start by analyzing the 
non-interacting case, i.e. $U_i=0$, which can be solved exactly.
We restrict our calculations to the conductance through the QDs 
coupled to the normal leads, therefore, 
we can simplify further Eq.~\eqref{eqcurr}, yielding $G_i=G_{cps}+G_{lar;i}+G_{qp;i}$,
where
\begin{align}
&G_{cps}=2\gamma_1\gamma_2 \int d \omega (f'_{e_1}(\omega)-f'_{h_2}(\omega))|\widehat{G}^a_{e_1,h_2}|^2 
\label{gcps}\\
&G_{lar;i}=4\gamma_i^2\int d \omega (f'_{e_i}(\omega)-f'_{h_i}(\omega))|\widehat{G}^a_{e_i,h_i}|^2\\
&G_{qp;i}=2\gamma_i  \int d \omega   \text{Im}\{ g_m\}
(f'_{e_i}(\omega)(|\widehat{G}^a_{e_i,e_m}|^2
+|\widehat{G}^a_{e_i,h_m}|^2)
\nonumber \\
&-f'_{h_i}(\omega)(|\widehat{G}_{h_i,h_m}|^2+|\widehat{G}_{h_i,e_m}|^2)),
\end{align}
are expressed in units of $G_0=2e^2/h$.
The interacting limit contains extra contributions 
coming from taking into account 
the second term of Eq.~\eqref{eqgfgl}.

From the previous expressions, we can define and calculate
the efficiency of the CPS, given by
\begin{align}
\chi=\frac{2G_{cps}}{G_1+G_2}.
\end{align}
It is worth mentioning that the corrections entering in the interacting limit come from relaxation processes. 
Thus, $\chi$ can be calculated exactly in both limits because the extra 
contributions enter only in $G_1$ and $G_2$ and not in $G_{cps}$.

\begin{figure}[t]
\centering
\includegraphics[width=3.5in,clip]{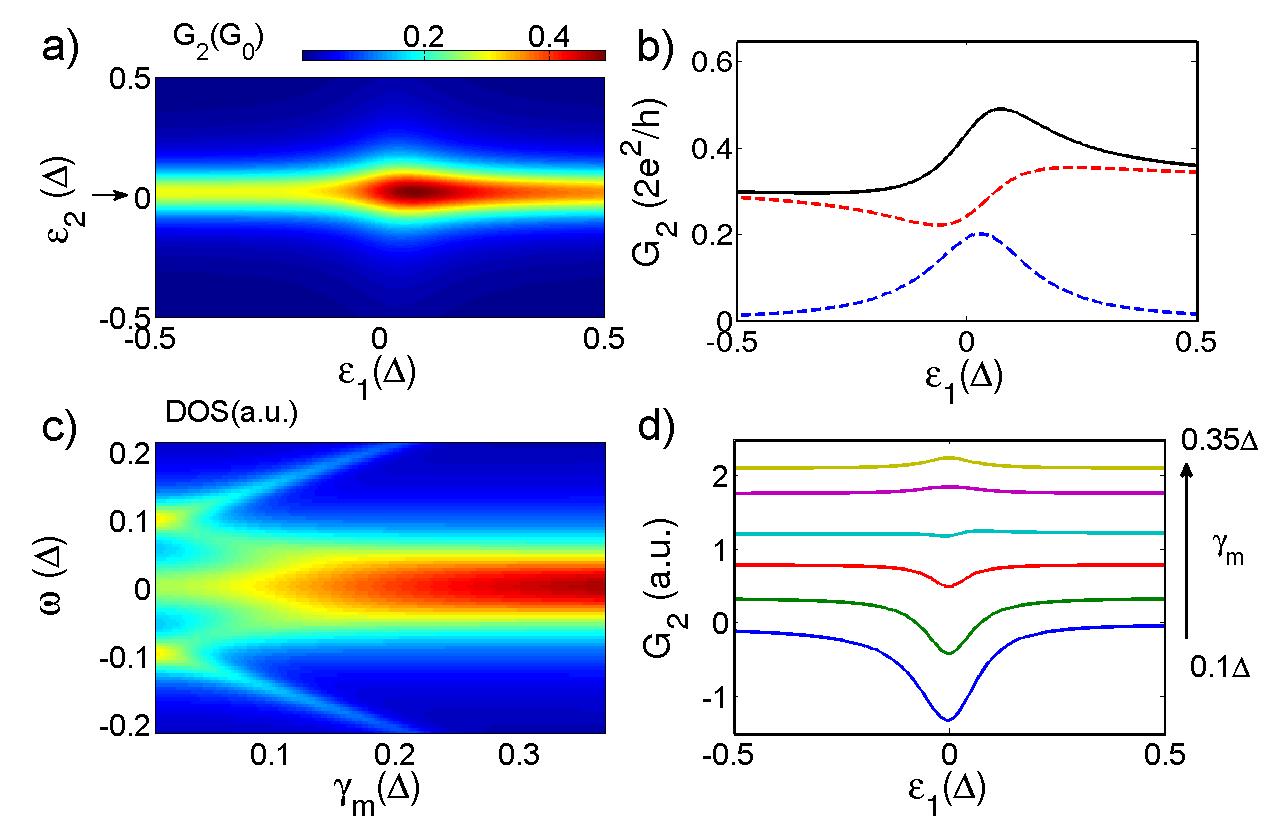}
\caption{(Color online) Panel a: Conductance G$_2$ vs $\epsilon_1$ and $\epsilon_2$, for the model parameters
$\gamma_1=\gamma_2=0.05\Delta$, $\epsilon_m=0.1\Delta$, t$_{1m}$=t$_{2m}=0.07\Delta$ and $\gamma_m=0.1\Delta$.
Panel b: Shows a cut along the resonance of panel~a, at the value $\epsilon_2$ marked by the arrow. We show the total conductance (solid black curve), 
and the decomposition into local (dashed red curve) and CPS contributions (dashed blue curve). 
Panel c: shows the DOS as a function of $\gamma_m$ and $\omega$, for the same model parameters used in panels~a and~b, but
with $\epsilon_m=0$.
Panel d: Shows cuts along the resonance for increasing values of $\gamma_m$ 
from bottom to top (from $\gamma_m=0.1\Delta$ up to 0.35$\Delta$). Here, we used 
same parameters as in panel c. All these plots were calculated using the non-interacting Hamiltonian, i.e.~$U_i=0$.
}
\label{Gnon-int}
\end{figure}

\section{Linear conductance} 

In this section we present the main general features of the linear regime. 
We will restrict our discussion to the non-interacting case because in the 
absence of an external magnetic field the interacting and the non-interacting cases exhibit a very similar behavior, 
except for some quantitative differences.
We will explore two different regimes: $|\epsilon_m|\gtrsim t_{im}$ in which coherent effects are less dominant, 
and $|\epsilon_m| \ll t_{im}$ for which coherent effects are dominant.

$|\epsilon_m| \gtrsim t_{im}$--- In this regime the conductance 
as a function of the dot levels $\epsilon_1$ and $\epsilon_2$ exhibits 
typically the shape shown in Fig.~\ref{Gnon-int}~a, independently of $\gamma_{1,2}$ and $\gamma_m$. 
In order to study the influence of the 
opposite QD levels on the conductance, we make a cut along the resonance. 
The resulting cut, $G_2$ vs $\epsilon_1$, is shown in Fig.~\ref{Gnon-int}~b, decomposed into
the CPS and the local contributions (see dashed lines in Fig.~\ref{Gnon-int}~b). 
We observe 
that the local contribution (dashed red line) exhibits an asymmetric Fano-shape, while the 
CPS contribution is characterized by a Lorentzian shape (dashed blue line).
This effect has been observed experimentally and explained theoretically in a recent publication [\onlinecite{Fulop2015a}]. 
The reason for the different behavior of the CPS and local contributions can be
traced to the effect of higher order tunneling processes in each case. 
In the case of any local process (AR or qp), the first finite contribution to the non-local response
involves
a destructive interference between two different paths,
i.e.~either particles go directly to the corresponding lead, or they give a detour 
through the opposite dot and then comes back to the corresponding dot (see Fig.~\ref{Fig1}~c). 
In the case of the CPS this interference is not present at the lowest in tunneling and thus 
it exhibits essentially a symmetric Lorentzian shape (see Fig.~\ref{Fig1}~b). 

In the limit of $\epsilon_m \gg \gamma_{m},~t_{im}$ we can expand 
to lowest order the GF, and obtain
the corresponding conductances at $T=0$ ($\omega=0$),
\begin{align}
&G_{cps}\sim \frac{4\gamma_1\gamma_2  t_{1m} t_{2m}\gamma_{m}^2}{\epsilon_m^2(\epsilon_1^2+\gamma_1^2)(\epsilon_2^2+\gamma_2^2)}, \\
&G_{lar;i} \sim \frac{8 t_{im}^4\gamma_i^2 \gamma_m^2 }{\epsilon_m^4(\epsilon_i^2+\gamma_i^2)^2}  \left(1+\frac{2\epsilon_j  t_{jm}^2}{\epsilon_m(\epsilon_j^2+\gamma_j^2)}\right)^2,
\label{eqlar}
\\
&G_{qp;i}\sim \frac{4 t_{im}^2 \gamma_i \text{Im}\{ g_m\}}{\epsilon_m^2(\epsilon_i^2+\gamma_i^2)}  \left(1+\frac{ \epsilon_j t_{jm}^2}{\epsilon_m(\epsilon_j^2+\gamma_j^2)}\right)^2,
\end{align}
expressed in units of $G_0=2e^2/h$.
Here the {\it j}-index stands for the dot in opposite position 
to {\it i} and the conductances are given in units of $2e^2/h$.
From these expressions we can see that the CPS contribution 
exhibits a Lorentzian shape, while the 
LAR and qp contributions an asymmetric Fano shape. 
From the LAR and qp expressions, it is easy to see that the 
asymmetry orientation depends on the sign of $\epsilon_m$. 
In addition, the asymmetry strength depends on the absolute 
value of $\epsilon_m/t_{im}$. 
Therefore, one can relate the energy $\epsilon_m$ to the shape-parameter 
that appears in every Fano-like curve [\onlinecite{Fano1961a}]. 
Interestingly, the $G_{lar;i}$ and the $G_{qp;i}$ contributions exhibit 
the same dependence except for a factor 2 in the interference term. This factor
accounts for the two particles that are involved in the local Andreev process.
This extra factor resuts in a more 
asymmetric pattern with more pronounced dips and the dips more pronounced than the quasiparticle one.

\begin{figure}[t]
\centering
\includegraphics[width=3.55in,clip]{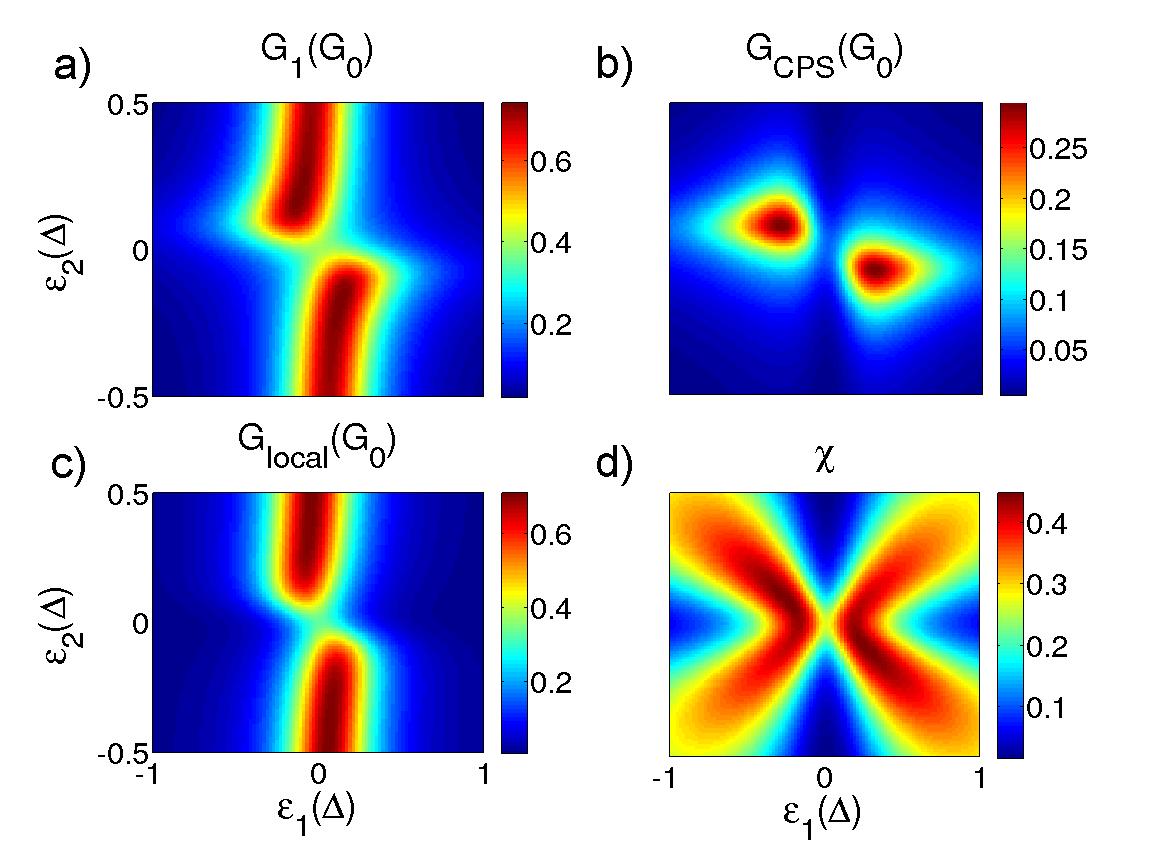}
\caption{(Color online) Total conductance (panel a), CPS contribution (panel b), local contribution (panel c) and efficency (panel c) 
as a function of $\epsilon_1$ and $\epsilon_2$ for the case of 
$\epsilon_m=0$, $t_{1m}=2t_{2m}=0.2\Delta$, $\gamma_1=\gamma_2=0.1\Delta$ and $\gamma_m=0.1\Delta$.}
\label{figcar}
\end{figure}

$|\epsilon_m| \ll t_{im}$--- In this limit the coherence effects are more important and 
the total conductance resonance shape may exhibit either a dip-shape or a peak-shape depending on 
the relation between $t_{im}$ and $\gamma_{1,m,2}$. 
The dip-shape resonance is produced by the presence of a dark state, i.e.~an eigenstate of the system 
which is not coupled to the central lead. 
In the limit of $\epsilon_1=\epsilon_2=0$ and $t_{1m}=t_{2m}$, the dark state is given by
\begin{align}
&|\Psi_{d}^{(e)}\rangle= \frac{1}{\sqrt{2}}\left(|1_e\rangle-|2_e\rangle \right),
\label{dark1}\\
&|\Psi_{d}^{(h)}\rangle= \frac{1}{\sqrt{2}}\left(|1_h\rangle-|2_h\rangle \right),
\label{dark2}
\end{align}
and it is placed at zero energy, for a very similar setup see Ref.~[\onlinecite{Michaelis2006a}]. 
Note that the presence of the dark state does not rely on the condition $t_{1m}=t_{2m}$. For $t_{1m}\neq t_{2m}$
the amplitudes of Eqs.~\eqref{dark1} and~\eqref{dark2} change as a function of $t_{1m}/t_{2m}$, but the behavior is qualitatively the same.
\begin{figure}[t]
\centering
\includegraphics[width=3.4in,clip]{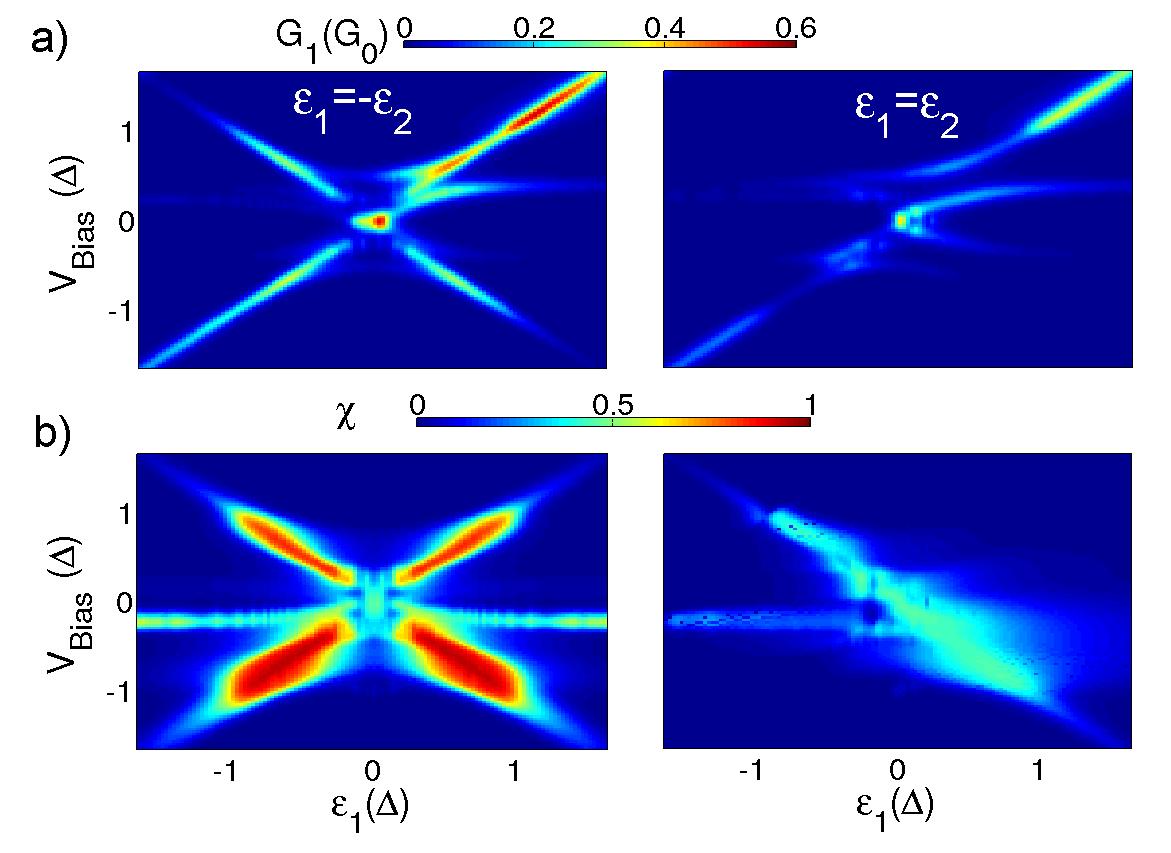}
\caption{(Color online) Conductance (panel a), and efficency (panel b) vs $\epsilon_1$ and $\mu$ 
for the antisymmetric case $\epsilon_1=-\epsilon_2$ (left column) and for the 
symmetric case $\epsilon_1=\epsilon_2$ (right column).
The parameters used are $\gamma_m=0.5\Delta$ and 
$t_{im}=0.21\Delta$, $\gamma_{1,2}=0.04\Delta$ and $\epsilon_m=0.3\Delta$
}
\label{Figurebias}
\end{figure}
For $\epsilon_m\ll t_{im}$ the closest excited conducting states are placed
at the energies 
\begin{align}
E_{exc}=\pm \frac{1}{\sqrt{2}} \sqrt{4t_{1m}^2+\gamma_m^2-\gamma_m\sqrt{8t_{1m}^2+\gamma_m^2}}.
\end{align}
Thus, the effect of increasing further $\gamma_{m}$ is to renormalize the higher conducting 
states towards the Fermi energy. 
This is illustrated in Fig.~\ref{Gnon-int}~c, showing the density of states
as a function of $\gamma_m$. 
As can be observed, the initially excited states  
converge towards the Fermi energy as long as we increase $\gamma_m$.
In addition, we can compare the evolution of the conductance resonance as we increase $\gamma_m$, 
from 0.1$\Delta$ to 0.35$\Delta$, shown in Fig.~\ref{Gnon-int}~d, with the corresponding density 
of states, shown in Fig.~\ref{Gnon-int}~c.
We observe that as long as the higher states get closer to the Fermi energy, the resonance turns from a dip into a peak. 
The value of $\gamma_m$ 
at which this occurs depends on the relative value of $t_{im}$ and $\gamma_{1,2}$.

It is also interesting to study the conductance decomposition for 
$\epsilon_m=0$. Here the resonance dip exhibits a width of the order of $t_{im}$. In this region, the transport 
is mainly blocked by the presence of the dark state.
In this situation, the CPS resonance splits into two peaks, and is the responsible for conferring 
some bending to the dip-shape resonance (compare Figs.~\ref{figcar}~a,~b and c). 
Thus, we see that the efficency exhibits a maximum value along the antidiagonal in the $\epsilon_1-\epsilon_2$-plane (see Fig.~\ref{figcar}~d).
It is worth mentioning that similar shapes have been already measured in Ref.~[\onlinecite{Fulop2015a}] when $\epsilon_m\approx 0$.
However, experimentally the conductance is not completely blocked as in the calculations. 
This could be explained in terms of transport activated by dephasing, produced by a fluctuating 
electromagnetic environment [\onlinecite{Marquardt2003a, Dominguez2011a, Contreras2014a}].
The reason for the splitting of the CPS contribution is the presence of an anticrossing in the central region spectral density.
When $\epsilon_m=0$, the states of QD~1 and~2 hybridize at the antidiagonal $\epsilon_1=-\epsilon_2 \sim t_{im}$, and thus provokes
the maximum values of the CPS conductance.

\begin{figure}[t]
\centering
\includegraphics[width=3.4in,clip]{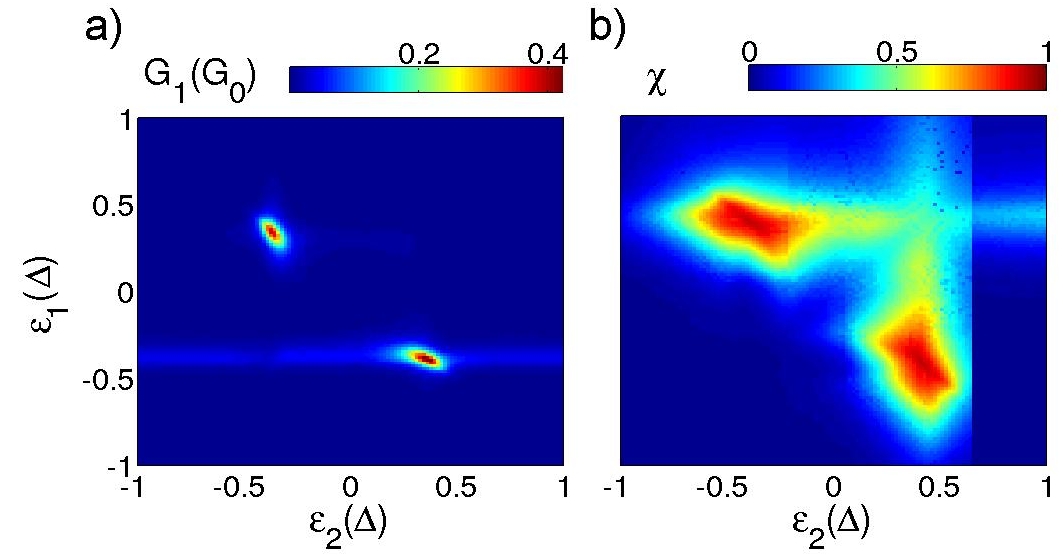}
\caption{(Panel~a) Conductance and (panel~b) efficency as a function of $\epsilon_1$ and $\epsilon_2$ 
for a fixed bias voltage $\mu=-0.65\Delta$ 
The rest of the parameters are the same as in Fig.~\ref{Figurebias}.}
\label{Figurebiasepsilon}
\end{figure}

\section{Non-linear conductance} 

In this section we present results for the conductances at finite bias in the
presence of Coulomb interactions.
In contrast to the linear regime, here we restrict to the interacting case 
because there are qualitative differences between both regimes. In addition,
we would like to compare the results from the 3-sites model with
the primary results obtained in the presence of interactions on Ref.~[\onlinecite{Recher2001a}].
One of the main results of the former work is that for 
$\Delta, U, |\epsilon_1-\epsilon_2| > \mu > \gamma_i$, with $\gamma_m<\gamma_{1,2}$
the CPS conductance is proportional to
\begin{align}
G_{CPS}\propto \frac{1}{((\epsilon_1-\mu)^2+\gamma_1^2)((\epsilon_2+\mu)^2+\gamma_2^2)},
\label{patrik}
\end{align}
and thus, exhibits maximum values at $\epsilon_1=-\epsilon_2$. 
The physical idea behind this behavior is
energy conservation. 
Initially, the Cooper pair starts from the superconducting lead, and thus,
$E_{i}=\epsilon_{F}=0$. 
Once the Cooper pair splits onto the normal leads, 
the final state is a superposition of two electrons placed at the normal leads, with energy 
$E_f=\epsilon_p+\epsilon_q$, where $\epsilon_{p,q}$ are the energies of the 
transmitted electrons to the normal leads. 
Energy conservation requires that $\epsilon_p=-\epsilon_q$, 
which leads to the sign difference of $\pm \mu$, observed 
in the denominator of Eq.~\eqref{patrik}.

Although the condition for observing a maximum CPS at 
$\epsilon_1=-\epsilon_2$ for $\Delta, U, |\epsilon_1-\epsilon_2| > \mu > \gamma_i$ 
is rather general, 
it is not clear whether it would hold for the 3-sites model.
Mainly, because the eigenenergies of the system are not only set 
by $\epsilon_{1,2}$, but also by $t_{im}$, $\gamma_m$ and $\epsilon_m$.
Nevertheless, we can observe in Fig.~\ref{Figurebias}~b 
that for $\mu>\gamma_m$, the antisymmetric configuration (left column) exhibits 
clearly an enhanced efficency with respect to the symmetric configuration (right column). 
In this figure we also observe other interesting features. For $|\mu|<\gamma_m$ we observe 
the contribution of Andreev bound states and for $|\mu| > \Delta$, the efficency 
drops due to an enhancement of the quasiparticle contribution.

In Fig.~\ref{Figurebiasepsilon} 
we set a fixed value of the bias voltage, i.e.~$\mu=-0.65\Delta$, 
where $\chi\sim 1$ for $\epsilon_1\sim -\epsilon_2$,
and then represent the conductance (panel~a) and the efficency (panel~b) 
in the $\epsilon_1$-$\epsilon_2$ plane.
We have obtained maximum efficiency values close to unity, and
observed numerically that $\chi$ decreases when $\gamma_{1,2} \gtrsim \gamma_m$ (not shown).
In general, we observe that $\chi$ becomes enhanced for $\gamma_m > t_{im}>\gamma_{1,2}$.
Note that this contradicts the previous results obtained by Recher et. al~[\onlinecite{Recher2001a}].
This occurs because in that work the condition $\gamma_m<\gamma_{1,2}$ was assumed to avoid overlap between the transmitted Cooper 
pairs.
Here, the pairing correlations are induced in the central part, 
and thus, they depend on the coupling parameters $t_{im},~\gamma_m$ 
and $\epsilon_m$.

\begin{figure}[t]
\centering
\includegraphics[width=2.7in,clip]{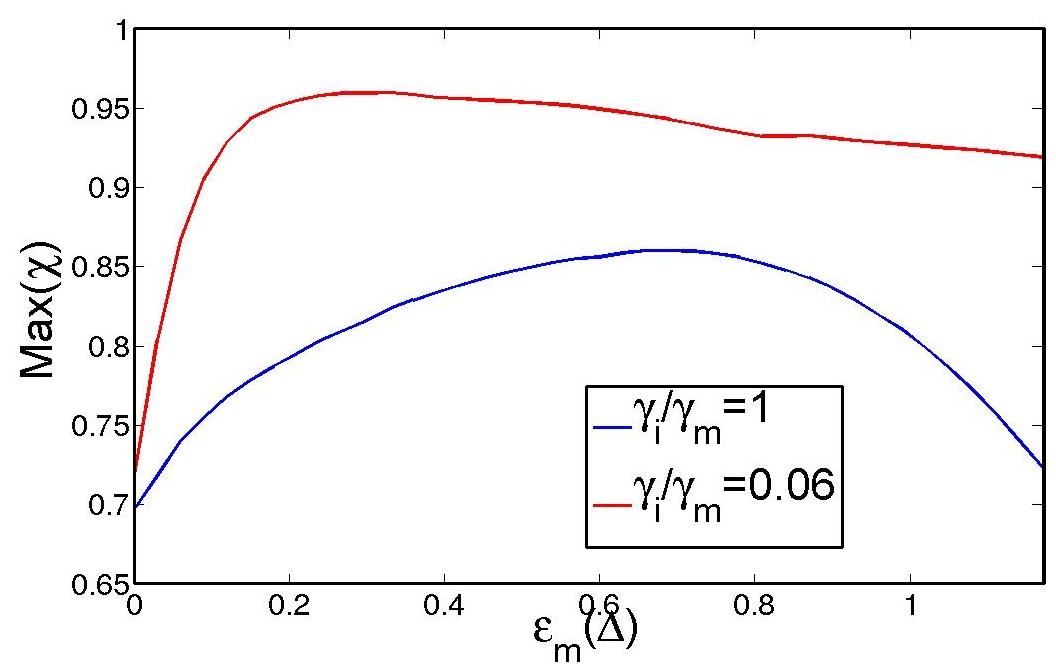}
\caption{(Color online) Maximal efficency obtained from the density plots $\chi$ vs $\epsilon_1=-\epsilon_2$ and
$\mu$, as a function of $\epsilon_m$ for two different cases $\gamma_i/\gamma_{m}=0.06$ 
(red curve) and $\gamma_i/\gamma_{m}=1.0$ (blue curve). 
We used $t_{im}=0.125\Delta$, $\gamma_m=0.3\Delta$.}
\label{effem}
\end{figure}

In Fig.~\ref{effem} we plot the maximum value of the efficency for a 
biased antisymmetric configuration ($\epsilon_1=-\epsilon_2$) as a function of $\epsilon_m$.
As we can observe, 
$\epsilon_m$ also plays an important role in the tuning of the efficency, especially for $\gamma_m\sim \gamma_{1,2}$. 
When $\gamma_m\gg \gamma_{1,2}$, the maximum of the efficency is rather flat for $\epsilon_m\gtrsim t_{im}$. 
On the other hand, when $\gamma_{1,2} > \gamma_m$, the maximum efficiency deviates from units and exhibits a maximum for $\epsilon_m \sim \gamma_m$.

In summary, we require
\begin{align}
\Delta, U, >|\epsilon_1-\epsilon_2| > \mu > \epsilon_m, \gamma_{m} > t_{im}>\gamma_{1,2}
\end{align}
for reaching the maximal efficency.
The reason for this hierarchy, is the following: 
In first place, we need to fulfill the condition 
$\epsilon_1=-\epsilon_2\approx\mu$ coming from energy conservation. 
In addition, in order to maintain the local contribution as low as possible, 
$\mu$ must be greater than the induced gap $\gamma_m$. 
Then, due to the discrete nature of the central site, 
we need 
$t_{im}\gtrsim |\epsilon_i-\epsilon_m|$ and $\gamma_m$ 
needs to be of the order of $|\epsilon_m|$, 
to have a significant effective electron-hole coupling. 

\section{conclusions}

In summary, we have studied the linear and non-linear conductance of the 3-sites model. 
In the linear regime we have analyzed the numerical results for two different regimes of $\epsilon_m$: 
$|\epsilon_m|>t_{im}$, and $|\epsilon_m|\ll t_{im}$.
For $|\epsilon_m|>t_{im}$, we obtained asymmetric resonances that can be decomposed into local processes 
and CPS processes, which exhibit Fano-resonant shapes and Lorentzian-resonance shapes, respectively. 
In order to understand the shape of the resonances, we describe the mechanism of the processes and 
provide some analytical expressions.
For $|\epsilon_m|\ll t_{mi}$, the coherence effects are stronger, and for certain conditions, 
the total conductance resonances exhibit pronounced dips. We show that these dips can be related to
the presence of a dark state that decouples the 3-sites from the superconducting lead.
As we have shown, in this limit the CPS conductance splits into two Lorentzian-resonances
placed at the antidiagonal in the $\epsilon_1$-$\epsilon_2$-plane.
In the finite bias limit, we have established a hierarchy of the 
model parameters that allows to optimize the CPS efficency.
Our results open the possibility for enhancing the efficency just by tunning 
the coupling parameters $t_{im}$, $\gamma_{1,2}$ and $\epsilon_{m}$. 
Thus, the new generation of Cooper pair splitters becomes a very
promising platform to produce and detect the entanglement in transport measurements.

\appendix 

\section{Advanced and Retarded Green's functions}

In this appendix we provide the exact expressions for the 
Green functions coupled to the leads in the non-interacting case, and the Lehmann representation 
for the uncoupled interacting limit.

\subsection{Non-interacting Hamiltonian: $U_i=0$}

Using the non-interacting Hamiltonian it is possible 
to express exactly the GFs accounting the coupling to the leads.
They have the form,
\begin{align}
\widehat{G}^{r/a}=(\hat{g}_0^{-1}(\omega) - \widehat{\Sigma}_0^{r/a}(\omega))^{-1},
\end{align}
where
\begin{equation}
\hat{g}_0(\omega)^{-1}= \begin{pmatrix}
                  \hat{s}_{e}            &     0    \\
                  0    &     \hat{s}_{h}\\
          \end{pmatrix} 
\end{equation}
with,
\begin{equation}
\hat{s}_{e/h}= \begin{pmatrix}
                  \omega \mp \epsilon_1     &     \mp t_{1m}      &    0   \\
                  \mp t_{1m}     &     \omega \mp \epsilon_m   &     \mp t_{2m}              \\
		     0         &        \mp t_{2m}    &      \omega \mp \epsilon_2      \\
          \end{pmatrix}
\end{equation}
is the Green function of the decoupled 3-sites. The self-energy that arise due to the 
coupling to the leads is given by
\begin{equation}
\Sigma^{r/a}(\omega)= \begin{pmatrix}
                         \hat{\lambda}^{r/a}     &     \hat{\Delta}^{r/a}    \\
                  \hat{\Delta}^{r/a}     &   \hat{\lambda}^{r/a}  \\
          \end{pmatrix}
\end{equation}
with 
\begin{equation}
\hat{\lambda}^{r/a}(\omega)= 
\begin{pmatrix}
                  \mp {\it i}\gamma_1     &  0   &    0\\
                  0     &     g_m^{r/a}(\omega)   &    0\\
		   0    &    0                &    \mp{\it i}\gamma_2  \\
\end{pmatrix},
\end{equation}
\begin{equation}
\hat{\Delta}^{r/a}(\omega)= 
\begin{pmatrix}
                  0     &  0   &    0\\
                  0     &  -f_m^{r/a}(\omega)  &    0\\
		   0    &    0                &    0 \\
\end{pmatrix}
\end{equation}
and
\begin{align}
&g_m^{r/a}(\omega)=-\gamma_{m}\frac{\omega \pm {\it i}\eta_s}{\sqrt{\Delta^2 -(\omega\pm {\it i} \eta_s)^2}}\\
&f_m^{r/a}(\omega)=\gamma_{m}\frac{\Delta}{\sqrt{\Delta^2 -(\omega\pm{\it i} \eta_s)^2}}.
\end{align}
The first term is the self-energy contribution coming from the quasiparticles present in the superconducting lead.
While the second term is the anomalous self-energy that induces a superconducting gap on the central site.

\subsection{Uncoupled Green's function of the interacting Hamiltonian: The Lehmann representation}

A simple approach to take into account interactions consists in
starting from the Hamiltonian of Eq.~\eqref{eq-H3s} without coupling to the leads, 
with a Hilbert space dimension of $4^3=64$. 
Then, a direct numerical diagonalization yields the states
$|\Phi_n\rangle$ with energies $E_n$. 
One can then use this basis to determine the uncoupled Green functions 
$\hat{g}_0^{r/a}(t,t') =
-i\theta(t-t') \langle\left[\psi(t),\psi^{\dagger}(t')\right]_+\rangle_0$
in terms of the multi-spinors.

Using the Lehmann representation $\hat{g}_0^{r/a}(\omega)$ can be written as
\begin{align}
\hat{g}_0^{r/a}(\omega) = \sum_{n,m}\frac{\left(e^{-\beta E_n} + e^{-\beta E_m}\right)}{Z} \frac{\langle \Phi_n | \psi |\Phi_m\rangle
\langle \Phi_m | \psi^{\dagger} |\Phi_n\rangle}{\omega - \left(E_m - E_n\right) \pm i0^+} , 
\end{align}
where $Z= \sum_n e^{-\beta E_n}$ and $\beta=1/k_BT$.

{\it Acknowledgments}---
We acknowledge S. Csonka and A. Baumgartner for enlightening discussions.
We gratefully acknowledge the financial support by the EU FP7 project SE$^2$ND, the EU 
ERC project CooPairEnt.

\end{document}